# Design triple points, nexus points and related topological phases by stacking monolayers

Yuee Xie, Cheng Gong, Jun Zhou, Yuanping Chen*

[1] *Faculty of Science, Jiangsu University, Zhenjiang, 212013, Jiangsu, China*

[2] *Department of Physics, Xiangtan University, Xiangtan, 411105, Hunan, China*

## Abstract

Triple points and nexus points are two interesting topological phases, which have been reported in some three-dimensional (3D) materials. Here, we propose that triple points, nexus points and related phases, such as topological tangle nodal lines, can be obtained by alternatively stacking two types of monolayers. Two conditions for the stacking monolayers are required: the first condition is that they have a three-fold ($C_3$) rotation symmetry and three mirror planes along the $C_3$ axis; the second condition is that one of the monolayers should be insulating while the other one should be metallic (or semiconducting) and has a double degenerate band and a nondegenerate band at Γ point around the Fermi level. Hexagonal boron nitride (HBN) and α/α′-boron sheets (α/α′-BS) are suggested as candidate materials. Even if HBN is a wide-gap insulator, the interactions between layers lead to crossings of the nondegenerate and double degenerate bands along the direction normal to the nanosheets, and thus form triple/nexus points or related phases. A tight-binding model is adopted to explain the phase transition between triple points, nexus points and other related phases.

Corresponding author: chenyp@ujs.edu.cn.

## I. Introduction

Topological materials, including topological insulators (TIs) and topological metals/semimetals (TMs)[1-6], have attracted much attentions recently because of new physics and potential applications behind them. Comparing with TIs, TMs possess various Fermi surfaces, such as zero-dimensional (0D) points[7-12], one-dimensional (1D) lines[13-18] and two-dimensional (2D) surfaces[19-21], and thus have tremendous classifications. These Fermi surfaces are formed by crossings of valence and conduction bands, protected by some specific symmetries. Weyl and Dirac points are two typical cases for the 0D nodal points on the Fermi surface[7-9], which correspond to two- and four-fold band crossing points, respectively, and reveal Dirac and Weyl fermions.

Triple point, formed by crossing of three bands, is another type of nodal point[22-32]. It can be further classified to two subtypes. One type is isolated nodal points like Weyl and Dirac points[22-24], the other type is two triple points in the momentum space connected by a trivial nodal line (crossed by two quadratic bands)[25-28]. Figure 1(a) presents a schematic view of the second type, which is usually protected by high structural symmetries: a three-fold ($C_3$) rotation axis, three mirror planes along the rotation axis and a mirror plane normal to the axis. The trivial nodal line linking the two triple point is along the rotation axis. After the mirror plane normal to the axis is eliminated, the two triple points will transit to a pair of nexus points[33-36], and the trivial nodal line split to four topological nodal lines. One of the four nodal lines still lies on the rotation axis, while the other three reside on the three mirrors, respectively, as shown in Fig. 1(e). The four splitting nodal lines merge at the nexus points.

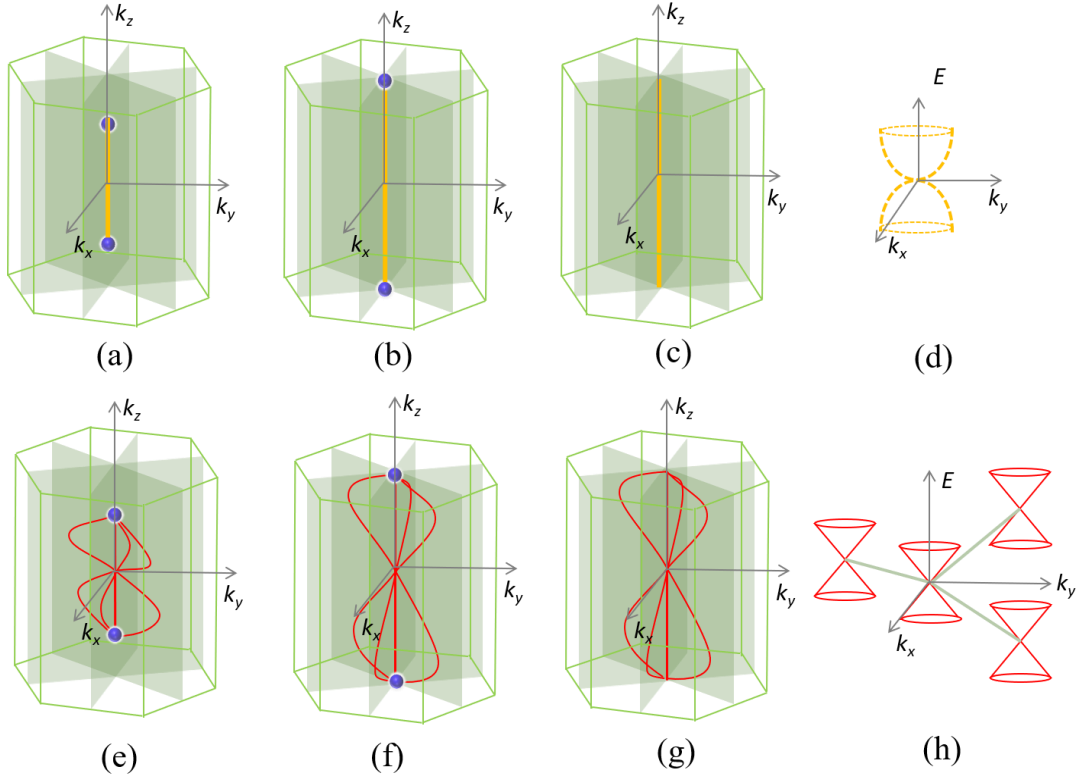

Figure 1. Topological phases in a first BZ. (a) A triple-points phase, where two (blue) triple points are connected by a yellow nodal line. (b) Similar to (a) but the triple points shift to the edge of the BZ. (c) A nodal line, which is a result of the triple points in (b) shifting out of the BZ. (e) A nexus-points phase, where two (blue) nexus points are connected by four nodal lines. (f) Similar to (e) but the nexus points shift to the edge of the BZ. (g) A phase of tangle nodal lines, which is a result of the nexus points in (f) shifting out of the BZ. The yellow nodal lines in (a-c) are trivial nodal lines, in which each point is a crossing of two quadratic bands as shown in (d). The red nodal lines in (e-g) are topological nodal lines, in which each point is a crossing of two linear bands as shown in (h). Each yellow nodal line in (a-c) split into four red nodal lines in (e-g), respectively.

In essence, the phase transition from triple points in Fig. 1(a) to nexus points in Fig. 1(e) is that the trivial (yellow) quadratic nodal line in the former phase splits to four (red) linear nodal lines in the latter[37,38], which is equal to the phase transition from a quadratic nodal point in Fig. 1(d) to four linear nodal points in Fig. 1(h). The triple-points and nexus-points phases can also evolve into other phases. For example, the triple points and nexus points move along the rotation axis to the edge of the first Brillouin Zone (BZ), and then two phases in Fig. 1(b) and 1(f) are formed, respectively. After the triple points or nexus points move out of the first BZ, only the nodal lines are remained. As a result, the triple-points phase transits to a quadratic nodal line, while the nexus-points phase transits to a phase of topological tangle nodal lines.

Besides the studies on topological phases and topological properties, searching new topological materials is another important aspect in topological field[39-42], for example, the triple-points and nexus-points phases have been proposed in some three-dimensional (3D) materials[42-45]. On the other hand, two-dimensional (2D) material has been another popular research field in recent decade[46-48]. These materials exhibit lots of excellent properties different from bulk materials, and thus have been applied in many fields[49,50], such as electronics, optoelectronics, thermal management, catalysis and energy. However, for the study of topological phases, 2D materials could be not a good choice. Although some topological phases have been found in 2D materials, such as Dirac point and nodal line, the dimensionality of BZ limits the existence of many other topological phases in 2D materials. For example, the triple points and nexus points mentioned above can only appear in a 3D BZ. In order to obtain these complicated topological phases based on 2D materials, one needs to extend the materials from 2D to 3D by stacking.

In this paper, we propose that triple points, nexus points and related topological phases in Fig. 1 can be realized by alternately stacking 2D insulator/metal layers. According to the symmetrical requirements of triple-points and nexus-points phases, we find that the stacking 2D materials should satisfy two conditions. Hexagonal boron nitride (HBN) and α/α′-boron sheets (α/α′-BS) are selected as candidates to confirm the idea. The calculations of first principles identify that there are triple points and tangle nodal lines in stacking HBN and α/α′-BS, respectively. The transitions between phases in Fig. 1 are explained by a tight-binding model based on the stacking structure.

## II. Two conditions for stacking monolayers

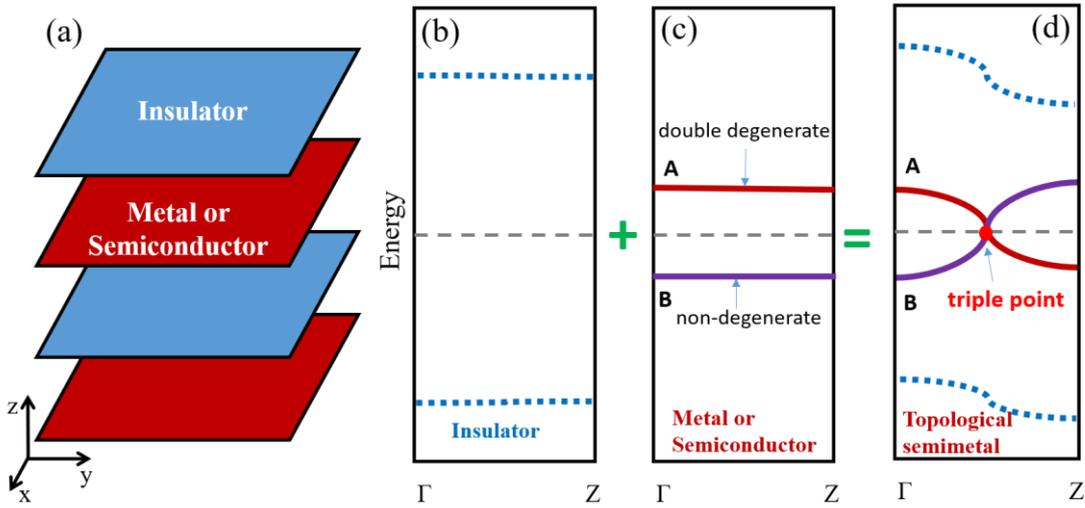

Figure 2. (a) A 3D layered structure formed by alternatively stacking monolayer insulators and metals/semiconductors. Schematic views of band structures along $\Gamma - Z$ for (b) a monolayer insulator, (c) a monolayer metal or a semiconductor, and (d) a 3D layered structure formed by alternatively stacking monolayers in (b) and (c). The red lines in (c-d) represent double degenerate bands, while the purple lines represent non-degenerate bands. The crossing point of the two bands in (d) is a triple point or nexus point.

To realize triple points, nexus points and related phases in Fig. 1, the 3D structures should have some certain symmetries mentioned above[25,33]. If the 3D structures are formed by stacking two types of monolayers, both of the two monolayers should have a $C_3$ rotation axis and three mirror planes normal to the layers. In this case, stacking structures can inherit these symmetry elements by aligning the $C_3$ axes of the two types of monolayers. Besides the symmetrical requirement, band structures of the monolayers should also meet some requirements. Triple points and nexus points are crossing points of a degenerate band along the $C_3$ rotation axis and a non-degenerate band. To generate the crossings, one can consider a stacking structure consisting of a metallic monolayer (or a narrow-gap semiconductor) and an insulating monolayer, as shown in Fig. 2(a). The insulating monolayer has a wide band gap (see Fig. 2(b)), while the metal (or narrow-gap semiconductor) should have a degenerate band and a non-degenerate band at Γ point around the Fermi level (see Fig. 2(c)). Because of absence of interlayer interactions before stacking, all the energy bands in Figs. 2(b-c) are exactly flat. After stacking, the interlayer interactions will bend the energy bands and may result in a triply-degenerate crossing point around the Fermi level, as shown in Fig. 2(d). If both of the monolayers exist a parallel mirror plane, triple points can be obtained. Otherwise, nexus points are obtained.

## III. Realization of topological phases in real materials

According to the former discussions, to generate triple points or nexus points, the two types of stacking monolayers at least meet two conditions: the first condition is that they have a $C_3$ rotation symmetry and three mirror planes along the $C_3$ axis; the second condition is that one of the monolayers should be insulating while the other one should be metallic (or semiconducting) and has a double degenerate band and a nondegenerate band at Γ point around the Fermi level. For the first condition, the nanosheets consisting of triangular or hexagonal rings are good candidates, such as hexagonal boron nitride, α/α′-BS, $C_3N$, etc. We select HBN and α/α′-BS as examples to form 3D stacking structures, because their energy bands satisfy the second conditions.

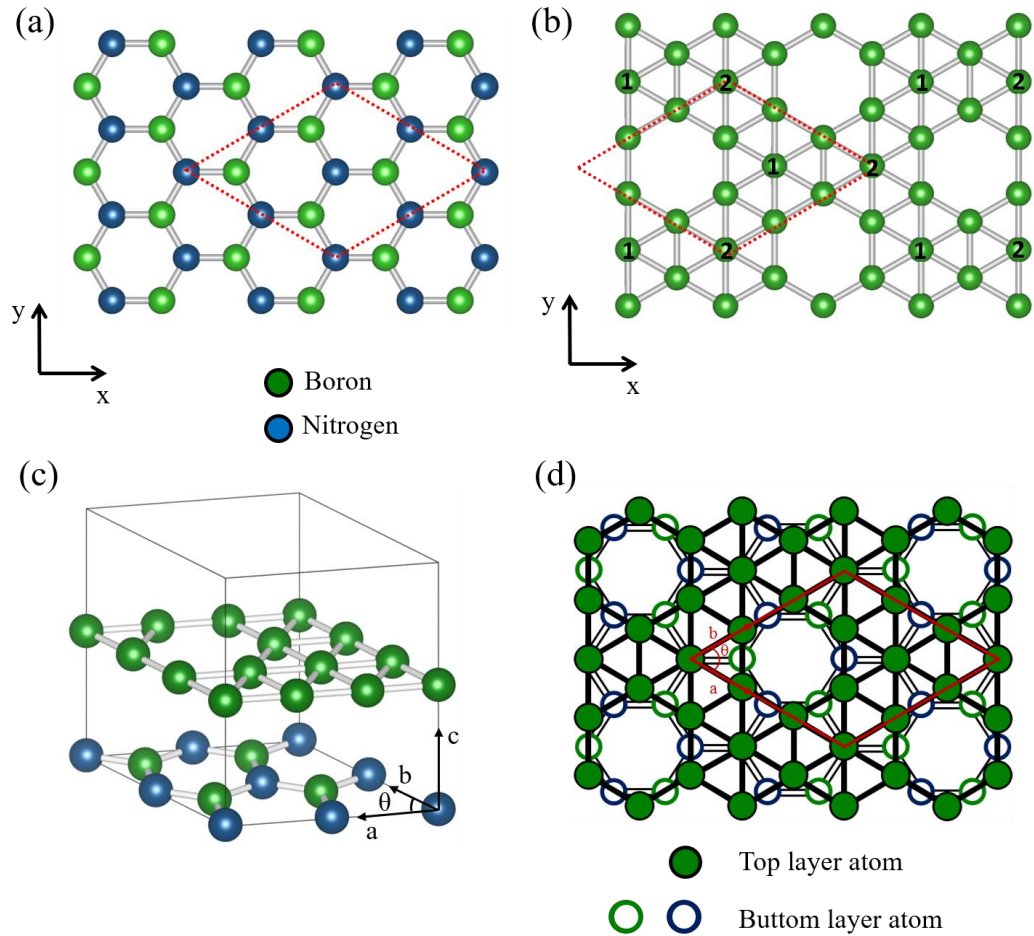

Figure 3. Atomic structure of (a) monolayer HBN and (b) monolayer α/α′-BS. α-BS is a perfect plane, while α′-BS is weak buckling because the atoms labeled 1 and 2 are inward and outward, respectively. (c) A primitive unit cell of a 3D structure formed by alternatively stacking HBN and α/α′-BS. The corresponding unit cells for the monolayer HBN and α/α′-BS are shown in the red dashed boxes in (a-b), respectively. (d) The top view of the 3D stacking structure in (c), whose primitive unit cell has been marked by red box.

Figures 3(a) and 3(b) show atomic structures of monolayer HBN and α/α′-BS, respectively. Both of them have a perpendicular $C_3$ axis and three mirror planes through the axis. HBN and α-BS have a mirror plane parallel to the layers because both of them are exactly flat[51]. However, α′-BS has no parallel mirror plane because the atoms labeled 1 and 2 in Fig. 3(b) are inward and outward[51], respectively. The lattice constants of primitive cells of α/α′-BS are $a_0 = b_0 = 5.21$ Å (the difference between α- and α′-BS is very small), and those of HBN are $a_1 = b_1 = 2.53$ Å. The mismatch between a 2×2 supercell of HBN and the primitive cell of α/α′-BS is smaller than 3%. Figures 3(c) and 3(d) show the side and top views of the 3D stacking structures, respectively. The optimized lattice constants of the stacking structures are $a_2 = b_2 = 2.53$ Å and $c_2 = 6.38$. It is noted that the stacking structure still has a C3 axis and three mirror planes along the axis.

Our first-principles calculations were based on the density functional theory within the PBE approximation[52] for the exchange-correlation energy. The core-valence interactions were described by projector augmented-wave (PAW)[53] potentials, as implemented in the VASP code[54]. Plane waves with a kinetic energy cutoff of 520 eV were used as the basis set. The calculations were carried out in primitive cells[55]. The atomic positions were optimized by using the conjugate gradient method, in which the energy convergence criterion between two consecutive steps was set at $10^{-5}$ eV. The maximum allowed force on the atoms is $10^{-2}$ eV Å$^{-1}$. The k-point meshes $7 \times 7 \times 6$, $15 \times 15 \times 1$ and $7 \times 7 \times 1$ were used for the Brillouin Zone (BZ) integration of 3D stacking structures, HBN and α/α′-SB respectively. When optimizing the atom positions of layered structures, the Grimme-D3 correction was included in the calculation to account for van der Waals interactions.

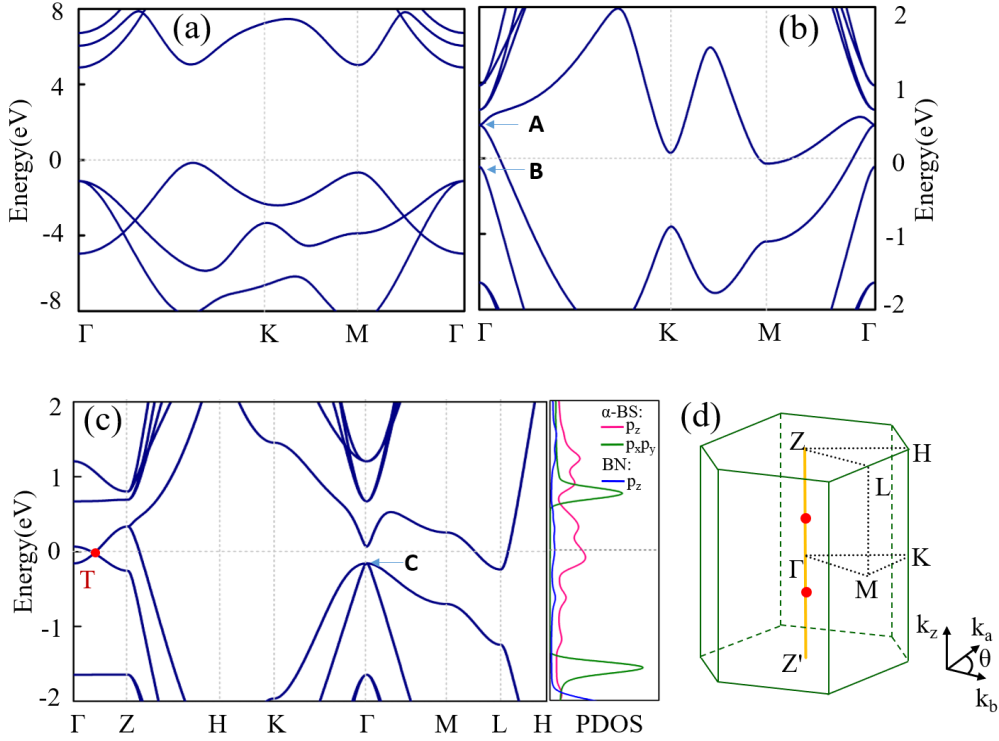

Figure 4. Band structures of (a) monolayer HBN and (b) monolayer α-SB. (c) Band structure and PDOS for the 3D stacking layered structure of HBN and α-SB. (d) Topological phase in the 3D stacking layered structure, where two red triple points are linked by a yellow nodal line along Γ-Z in the first BZ.

In Figs. 4(a-b), the band structures of monolayer HBN and $\alpha-$BSare shown, respectively. Monolayer HBN is a wide-gap insulator, whose band gap is larger than 4 eV. Monolayer $\alpha-$BS is a metal. There is a double degenerate point A and a nondegenerate point B at Γ around the Fermi level (see Fig. 4(b)). They will extend along $\Gamma-Z$ and then form flat bands like Fig. 2(c) in case of no interlayer coupling. Figure 4(c) shows the band structure for the 3D structure by alternatively stacking monolayer HBN and $\alpha-$BS. Due to interlayer interactions, the degenerate band extending from point A cross with the nondegenerate band extending from point B, and thus leads to a triple point at $k_z = 0.21$ $\pi/c$. There is another triple point locating at $-k_z$ because of time reversal symmetry. The degenerate band along $k_z$ axis is in fact a nodal line, and each point in the nodal line is a crossing point of two quadratic bands (see point C at Γ in Fig. 4(c)). Topological phase in the stacking strcture is shown in Fig. 4(d), where two red triple points are linked by a trivial nodal line. It also corresponds to the phase in Fig. 1(a).

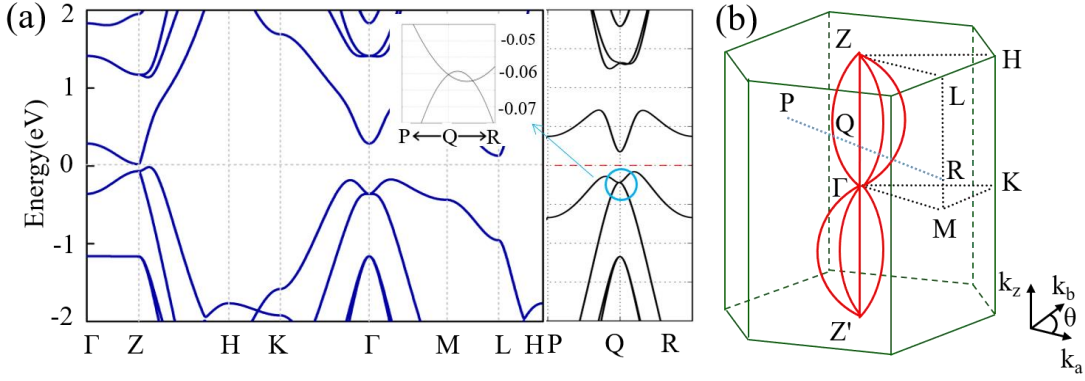

Figure 5. (a) Band structure for the 3D stacking layered structure of HBN and α'-SB. Inset: An amplified band structure in the light-blue circle region in the right panel. (b) Topological phase in the 3D stacking layered structure, where four nodal lines are tangled in the first BZ.

Comparing with monolayer α-BS, monolayer α′-BS also has a C3 rotation axis, but it has no parallel mirror plane. Figure 5(a) shows band structure for the alternative stacking HBN and α′-BS layers. One can find that, along Γ − Z, there is no crossing point between the degenerate and nondegenerate bands, and thus no triple points exist in momentum space. However, the degenerate band leads to a nodal line along $k_z$. A clear examination indicates that there are other nodal lines in momentum space. For example, we calculate band structure along a $k$ path P-Q-R on the mirror plane $k_b = 0$ (i.e. ΓZLM, see Fig. 5(b)), as shown in the right panel in Fig. 5(a), which illustrates that there are two crossing points between two bands (see the inset in Fig. 5(a)). One crossing point belongs to the straight nodal line along $k_z$, while the other belongs to another curved nodal line on the plane $k_b = 0$. Considering the $C_3$ rotation symmetry of the structure, on the other two mirrors, there also exist nodal lines. Figure 5(b) shows the topological phase in the first BZ of stacking HBN and $\alpha'$-BS layers. It is a phase of tangle nodal lines similar to that in Fig. 1(g).

## IV. Tight-binding model for the stacking structures

To explain the topological properties of the stacking structures, one can use a tight-binding (TB) model to describe these phases and phase transitions between them. From projected density of states (PDOS) in the right panel of Fig. 4(c), one can find that only $p_z$ orbitals of B, N atoms have contributions to the energy bands around the Fermi level, and thus only one $p_z$ orbital in these atoms need to be considered to describe electronic properties. If we consider one orbital of each atom in the primitive cell in Fig. 4(c), a TB model is found:

$$H = H_{BS} + H_{HBN} + H', \tag{1}$$

where $H_{BS}$, $H_{HBN}$, $H'$ represent Hamiltonians of monolayer $\alpha/\alpha' - BS$, HBN and their interlayer interactions, respectively, and

$$H_{BS} = \sum_i \varepsilon_1 a_i^\dagger a_i + \sum_{i,j} t_\alpha a_i^\dagger a_j, \tag{2}$$

$$H_{HBN} = \sum_l \varepsilon_2 b_l^\dagger b_l + \sum_m \varepsilon_3 b_m^\dagger b_m + \sum_{l,m} t_\beta b_l^\dagger b_m, \tag{3}$$

$$H' = \sum_{i,l} t_\gamma a_i^\dagger b_l, \tag{4}$$

where $a_i^\dagger/a_j, b_l^\dagger/b_m$ represent the creation/annihilation operators in monolayer $\alpha/\alpha' - BS$ and HBN, respectively, $\varepsilon_1$ represents site energy of B atom in $\alpha/\alpha' - BS$, $\varepsilon_2/\varepsilon_3$ represents site energy of B and N atoms in HBN, respectively, $t_\alpha$ and $t_\beta$ are intralayer hopping parameters in the $\alpha/\alpha' - BS$ and HBN, and $t_\gamma$ are interlayer hopping parameters between the monolayers (see Figs. 6(a-b)).

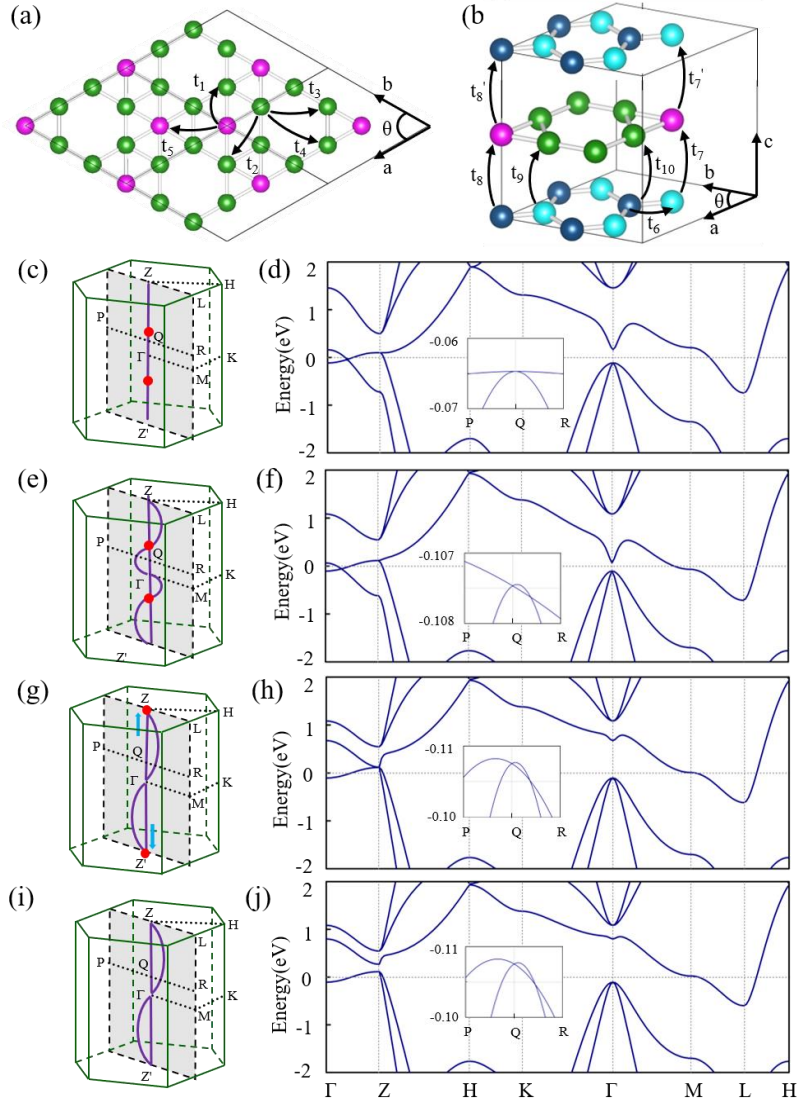

Figure 6. (a) Schematic views of the intralayer hopping parameters in monolayer α/α'-SB in Eq. (2). (b) Intralayer hopping parameter of HBN in Eq. (3) and interlayer hopping parameters between BN and α/α'-SB in Eq. (4). (c-j) Topological phases and corresponding band structures generated by Eq. (1) with different parameters, and the detailed values of the parameters are given in Table 1. (c) A triple-points phase similar to that in Fig. 1(a). (e) A nexus-points phase similar to that in Fig. 1(e). (g) A critical phase between phases of nexus points and tangle nodal lines. (i) A phase of tangle nodal lines. The topological phases in the left panels correspond to the band structures in the right panels, respectively. It is noted that, in (c), (g) and (i), only the nodal lines on one mirror plane are given to clearly show their locations.

By tuning the parameters in Eqs. (2-4), one can get all the phases in Fig. 1 (the detailed parameter values corresponding to the phases are given in Table 1). Figures 6(c-j) show four phases and their corresponding band structures. Figure 6(c) exhibits a triple-points phase, which is similar to that in Fig. 4(d) or 1(a). The corresponding band structure in Fig. 6(d) is also very similar to that in Fig. 4(c), indicating that the Eq. (1) can reproduce the band structure of stacking structure well. When the parallel mirror is eliminated, the interlayer parameters $t_\gamma$ should be changed, for example, $t_7 \neq t'_7$ and $t_8 \neq t'_8$. The variation of $t_\gamma$ leads to three different topological phases, as shown in Figs. 6(e), 6(g) and 6(i). The phase in Fig. 6(e) is a standard nexus-points phase similar to that in Fig. 1(e). Here, we only present the nodal line on one mirror plane. Figure 6(i) is a phase of tangle nodal lines, while Fig. 6(g) is a critical phase between nexus points and tangle nodal lines. From Fig. 6(c) to 6(i), the phase transition from triple points to nexus points to tangle nodal lines are clearly shown. The band structure in Fig. 6(f) seems very similar to that in Fig.6(d). However, the insets in the two figures illustrate their difference, which also indicate that the quadratic nodal line in the triple-points phase transits two linear nodal line on one mirror plane.

Table 1. Hopping energies in Eqs. (2-4) to generate topological phases in Figs. 6(c, e, g, i), and site energies are $\varepsilon_1 = -1.4$, $\varepsilon_2 = -3.5$ and $\varepsilon_3 = -3.0$. All values are in units of eV.

| Topological phase | $t_1$ | $t_2$ | $t_3$ | $t_4$ | $t_5$ | $t_6$ | $t_7$ | $t_7'$ | $t_8$ | $t_8'$ | $t_9$ | $t_{10}$ |
|---|---|---|---|---|---|---|---|---|---|---|---|---|
| Fig. 6(c) | -3.0 | -0.7 | -0.3 | -0.2 | -0.2 | -2.0 | 1.0 | 1.0 | 0.8 | 0.8 | 0.6 | 0.5 |
| Fig. 6(e) | | | | | -0.25 | | 1.0 | 0.5 | 0.8 | 0.4 | 0.6 | 0.5 |
| Fig. 6(g) | | | | | -0.37 | | 0.95 | 0.45 | 0.75 | 0.35 | 0.55 | 0.45 |
| Fig. 6(i) | | | | | -0.5 | | 0.9 | 0.4 | 0.7 | 0.3 | 0.5 | 0.4 |

## V. Conclusions

In summary, we have designed a method to produce triple points, nexus points or tangle nodal lines by alternately stacking monolayers. Two necessary conditions for the stacking monolayers are proposed. According to the conditions, HBN, α/α′-BS are selected as candidates. Even if the HBN is a wide-gap insulator, the interactions between layers of HBN and α-BS lead to triple points along the direction perpendicular to the layers. The stacking of HBN and α′-BS generates a phase of tangle nodal lines, because the parallel mirror plane is eliminated.

Some other monolayer materials satisfying the two conditions can also be used to generate triple points, nexus points and related phases. Other topological phases, such as nodal chain, Hopf link and nodal surface, should be also obtained by stacking different monolayers as long as certain symmetries are satisfied. Therefore, our work combines the two fields of 2D materials and topological phases together, and paves a way to search topological materials by stacking 2D monolayers.

## Acknowledgments

This work was supported by the National Natural Science Foundation of China (No. 11874314).